\documentclass[12pt]{article}
\usepackage[dvips]{color}
\usepackage{epsfig}
\usepackage{amsmath}
\usepackage{graphicx}
\def\Box{\hbox{$\rlap{$\sqcup$}\sqcap$}}
\begin{document}
\setcounter{page}{1}

\pagestyle{plain} \vspace{1cm}

\begin{center}
\Large{\bf {Phantom Divide Crossing with General Non-minimal Kinetic Coupling}}\\
\small \vspace{1cm} {\bf A. Banijamali $^{a}$
\footnote{a.banijamali@nit.ac.ir}} and {\bf B. Fazlpour $^{b}$
\footnote{b.fazlpour@umz.ac.ir}}\\
\vspace{0.5cm}  $^{a}$ {\it Department of Basic Sciences, Babol
University of Technology, Babol, Iran\\} \vspace{0.5cm} $^{b}$ {\it
Department of Physics, Ayatollah Amoli Branch, Islamic Azad
University, P. O. Box 678, Amol, Iran\\}
\end{center}
\vspace{1.5cm}
\begin{abstract}
We propose a model of dark energy consists of a single scalar field
with a general non-minimal kinetic couplings to itself and to the
curvature. We study the cosmological dynamics of the equation of
state in this setup. The coupling terms have the form $\xi_{1} R
f(\phi)\partial_{\mu}\phi\partial^{\mu}\phi$ and $\xi_{2}
R_{\mu\nu}f(\phi)\partial^{\mu}\phi\partial^{\nu}\phi$ where
$\xi_{1}$ and $\xi_{2}$ are coupling parameters and their dimensions
depend on the type of function $f(\phi)$. We obtain the conditions
required for phantom divide crossing and show numerically that a
cosmological model with general non-minimal derivative coupling to
the scalar and Ricci curvatures can realize such a crossing.\\

{\bf PACS numbers:} 95.36.+x, 98.80.-k, 04.50.kd\\
{\bf Keywords:} Crossing of phantom divide; Non-minimal derivative
coupling;
Tachyon field.\\

\end{abstract}
\newpage
\section{Introduction}
Recent observational data from CMB temperature fluctuations
spectrum, Supernova type Ia redshift-distance surveys and other data
sources, have shown that the universe is currently in a positively
accelerated phase of expansion [1-4]. Nevertheless, there is not
enough standard matter density in the universe to support this
accelerated expansion. Therefore, we need additional cosmological
component dubbed dark energy to explain this achievement. Dark
energy (DE) has been one of most active field in modern cosmology
[5]. The simplest candidate  for DE is a tiny positive
time-independent cosmological constant $\Lambda$. However, this
scenario suffers from some difficulties such as lack of physical
motivation, huge amount of fine-tuning to explain cosmological
accelerated expansion and no dynamics for its
equation of state [6].\\
As a possible solution to these problems, many dynamical scalar
field models of DE have been proposed. Quintessence, phantom,
k-essence and tachyon scalar fields belong to these sort of DE
models (for review see [6]).\\
In the other hand, there are some datasets (such as the Gold
dataset) that show a mild trend for crossing of the phantom divide
line by the effective equation of state (EoS) parameter $\omega$
(the ratio of the effective pressure of the universe to the
effective energy density of it) of dark component. The equation of
state parameter crosses the phantom divide line ($\omega=-1$) at
recent redshifts and current accelerated expansion
requires $\omega<-1/3$ .\\
The quintom scenario of dark energy is designed to understand the
nature of dark energy with $\omega$ across $-1$. To realize a viable
quintom scenario of dark energy it needs to introduce extra degree
of freedom to the conventional theory with a single fluid or a
single scalar field. The first model of quintom scenario of dark
energy is given by Ref. [7] with two scalar fields (quintessence and
phantom). Another attempts for constructing a quintom model are as
follows: scalar field model with non-linear kinetic terms [8] or a
non-linear higher-derivative one [9], braneworld models [10],
phantom coupled to dark matter with an appropriate coupling [11],
string inspired models [12], non-local gravity [13], modified
gravity models [14] and also non-minimally coupled scalar field
models in which scalar field couples with scalar curvature,
Gauss-Bonnet invariant or modified $f(R)$ gravity [15-17]. Crossing
of the phantom divide can also be realized with single imperfect
fluid [18] or by a constrained single degree of freedom dust like
fluids [19]. It has been shown in [45] that in the future the
crossing of the phantom divide are the generic feature for all the
existing viable $f(R)$ model such as Hu- Sawicki [46], Starobinsky
[47], Tsujikawa [48] and the exponential gravity [49, 50] models. A
phantom crossing DGP model has been constructed in [51] and the
interacting chaplygin gas dark energy model which realizes phantom
crossing investigated in [52] (for a detailed review on extended
theories of gravity and
their cosmological applications see [53]).\\
Furthermore, non-minimal couplings are generated by quantum
corrections to the scalar field theory and they are essential for
the renormalizability of the scalar field theory in curved space
[20, 21]. One can extend the non-minimally coupled scalar tensor
theories, allowing for non-minimal coupling between the derivatives
of the scalar fields and the curvature [22]. Such a non-minimal
coupling may appear in some Kaluza-Klein theories [23, 24] and also
as quantum corrections to Brans-Dicke theory [25]. A model with
non-minimal derivative coupling was proposed in [22, 26, 27] and
interesting cosmological behaviors of such a model in inflationary
cosmology [28], quintessence and phantom cosmology [29, 30],
asymptotic solutions and restrictions on the coupling parameter [31]
have been widely studied in the literature. General non-minimal
coupling of a scalar field and its kinetic term to the curvature as
a source of late-time cosmic acceleration has been analyzed in [32].
Also, non-minimal coupling of modified $f(R)$ gravity and kinetic
part of Lagrangian of a massless scalar field has been investigated
in [33, 41, 42]. It has been shown that inflation and late-time
cosmic acceleration of the universe can be realized in such a model.
In this paper we consider the function $f(R)$ as linear in $R$ but
we generalize the model allowing extra $R_{\mu\nu}$ coupling with
kinetic term of the scalar field. We are interested in our analysis
to the case of tachyon scalar field.\\ The tachyon field in the
world volume theory of the open string stretched between a D-brane
and an anti-D-brane or a non-BPS D-brane plays the role of scalar
field in the context of string theory [34]. What distinguishes the
tachyon Lagrangian from the standard Klein-Gordan form for scalar
field is that the tachyon action has a non-standard type namely,
Dirac-Born-Infeld form [35]. Moreover, the tachyon potential is
derived from string theory and should be satisfy some definite
properties to describe tachyon condensation and other requirements
in string theory. In summary, our motivation for investigating a
model with non-minimal derivative coupling and tachyon scalar field
is coming from a fundamental theory such as string/superstring
theory and it may provide a possible approach to
quantum gravity from a perturbative point of view [36-38].\\
An outline of the present work is as follows: In section 2 we
introduce a model of DE in which the tachyon field plays the role of
scalar field and the non-minimal couplings between scalar field, the
derivative of scalar field, the Ricci scalar and the Ricci tensor
are also present in the action. Then we derive field equations as
well as energy density and pressure of the model in order to study
the EoS parameter behavior in section 3. We obtain the conditions
required for $\omega$ crossing $-1$ and using numerical method, we
will show that the model can realize the
$\omega=-1$ crossing. Section 4 is devoted to our conclusions.\\

\section{Field Equations}
We start with the following action for tachyon field with general
non-minimal derivative couplings to the scalar and Ricci curvatures
and also with itself,
\begin{equation}
S=\int d^{4}x
\sqrt{-g}\Big[\frac{1}{2\kappa^{2}}R-V(\phi)\sqrt{1+g^{\mu\nu}
\partial_{\mu}\phi\partial_{\nu}\phi}-\frac{1}{2}\xi_{1}
R f(\phi)\partial_{\mu}\phi\partial^{\mu}\phi-\frac{1}{2}\xi_{2}
R_{\mu\nu}f(\phi)\partial^{\mu}\phi\partial^{\nu}\phi\Big],
\end{equation}
where $\kappa^{2} = 8\pi G = \frac{1}{M_{Pl}^{2}}$ while $G$ is a
bare gravitational constant and $M_{Pl}$ is a reduced Planck mass,
$V(\phi)$ is the tachyon potential which is bounded and reaching its
minimum asymptotically. $f(\phi)$ is a general function of the
tachyon field $\phi$ and $\xi_{1}$ and $\xi_{2}$ are coupling
parameters and their dimensions depend on the type of function
$f(\phi)$. Note that if we consider the following restriction on
parameters $\xi_{1}$ and $\xi_{2}$,
\begin{equation}
2\xi_{1}+\xi_{2}=0,
\end{equation}
then the last two terms in action (1) reduced to $\xi_{1}
f(\phi)G_{\mu\nu}\partial^{\mu}\phi\partial^{\nu}\phi$ which
corresponds to non-minimal derivative coupling of scalar field with
the Einstein tensor. The cosmological implications of such a theory
have been studied in Refs. [28-32].\\
The models of kind (1) with non-minimal coupling between derivatives
of a scalar field and curvature are the extension of scalar-tensor
theories and it is shown that these theories cannot be recasting
into the Einstein gravity form by a conformal transformation of the
metric [22]. A theory with the derivative coupling term $\xi_{2}
R_{\mu\nu} \partial^{\mu}\phi
\partial^{\nu}\phi$ has been
considered in [31] and constraints on the coupling parameter
$\xi_{2}$ have been obtained using precision tests of general
relativity. A general model containing two derivative coupling terms
$\xi_{1} R
\partial_{\mu}\phi
\partial^{\mu}\phi$ and $\xi_{2} R_{\mu\nu} \partial^{\mu}\phi
\partial^{\nu}\phi$, has been discussed in [26, 27]. It was shown
that the de Sitter spacetime
is an attractor solution of the model if $4\xi_{1}+\xi_{2}>0$. In here we study the model (1)
 without the restriction (2) on the coupling parameters.  \\
Varying the action (1) with respect to metric tensor $g_{\mu\nu}$,
leads to
\begin{equation}
G_{\mu\nu}=R_{\mu\nu}-\frac{1}{2}g_{\mu\nu}R=\kappa^{2}\big(T_{\mu\nu}+\xi_{1}
T'_{\mu\nu}+\xi_{2} T''_{\mu\nu}\big),
\end{equation}
where
\begin{equation}
T_{\mu\nu}=V(\phi)\Big(\frac{\nabla_{\mu}\phi
\nabla_{\nu}\phi}{u}-g_{\mu\nu}u\Big),
\end{equation}
\begin{equation}
T'_{\mu\nu}=R\big(\nabla_{\mu}\phi\nabla_{\nu}\phi\big)+G_{\mu\nu}\big(\nabla\phi\big)^{2}-
\nabla_{\mu}
\nabla_{\nu}\big(\nabla\phi\big)^{2}+g_{\mu\nu}\Box\big(\nabla\phi\big)^{2},
\end{equation}
and
$$T''_{\mu\nu}=-\frac{1}{2}g_{\mu\nu}\nabla_{\gamma}
\phi\nabla_{\lambda}\phi\,
R^{\gamma\lambda}+2\nabla_{\gamma}\phi\nabla_{(\mu}\phi
R^{\gamma}_{\nu)}+\frac{1}{2}\Box
(\nabla_{\mu}\phi\nabla_{\nu}\phi)$$
\begin{equation}
-\nabla_{\gamma}\nabla_{(\mu}(\nabla_{\nu)}\phi \nabla^{\gamma}\phi)
+\frac{1}{2}g_{\mu\nu}\nabla_{\gamma}\nabla_{\lambda}(\nabla^{\gamma}\phi\nabla^{\lambda}
\phi),
\end{equation}
here $u=\sqrt{1+\nabla_{\mu}\phi\nabla^{\mu}\phi}$.\\
One can obtain the scalar field equation of motion by variation of
action (1) with respect to $\phi$,
$$\nabla_{\mu}\Big(\frac{V(\phi)\nabla^{\mu}\phi}{u}\Big)-\frac{dV(\phi)}{d\phi}u+\frac{1}{2}
f(\phi)\nabla_{\mu}\big[\nabla_{\nu}\phi\big(\xi_{1}
g^{\mu\nu}R+\xi_{2} R^{\mu\nu}\big)\big]$$
\begin{equation}
-\frac{1}{2}\big[\xi_{1} R
\partial_{\mu}\phi\partial^{\mu}\phi+\xi_{2}
R_{\mu\nu}\partial^{\mu}\phi\partial^{\nu}\phi \big]
\frac{df(\phi)}{d\phi}=0.
\end{equation}
In a spatially-flat Friedmann-Robertson-Walker (FRW) spacetime with
the metric,
\begin{eqnarray}
ds^{2}=-dt^{2}+a^{2}(t)(dr^{2}+r^{2}d\Omega^{2}),
\end{eqnarray}
the  components of the Ricci tensor $R_{\mu\nu}$ and the Ricci
scalar $R$ are given by
\begin{equation}
R_{00}=-3\big(\dot{H}+H^{2}\big),\,\,R_{ij}=a^{2}(t)\big(\dot{H}+3H^{2}\big)\delta_{ij},\,\,
R=6\big(\dot{H}+2H^{2}\big),
\end{equation}
where $H=\frac{\dot{a}(t)}{a(t)}$ is the Hubble parameter and $a(t)$
is the scale factor. The equation (7) for a homogeneous
time-depending $\phi$ in FRW background (8) reads
$$\frac{\ddot{\phi}}{1-\dot{\phi}^{2}}+3H\dot{\phi}+\frac{1}{V(\phi)}
\frac{dV}{d\phi}+\frac{\sqrt{1-\dot{\phi}^{2}}}{V(\phi)}
\Bigg(\frac{3}{2}\Big[(2\xi_{1}+\xi_{2}) \dot{H}+(4\xi_{1}+\xi_{2})
H^{2}\Big]\Big(2f(\phi)\ddot{\phi}+\frac{df}{d\phi}\dot{\phi}^{2}\Big)$$
\begin{equation}
+9(4\xi_{1}+\xi_{2})
H^{3}f(\phi)\dot{\phi}+3(2\xi_{1}+\xi_{2})\ddot{H}f(\phi)\dot{\phi}+3(14\xi_{1}+5\xi_{2})
H \dot{H} f(\phi)\dot{\phi}\Bigg)=0.
\end{equation}
Using equations (4)-(6), the $(0,0)$ component and $(i,i)$
components of equation (3) correspond to energy density and pressure
respectively,
\begin{equation}
\rho=\frac{V(\phi)}{\sqrt{1-\dot{\phi}^{2}}}+9\xi_{1}
H^{2}f(\phi)\dot{\phi}^{2}+3(2\xi_{1}+\xi_{2})\Big(\dot{H}f(\phi)\dot{\phi}^{2}-H
f(\phi)\dot{\phi}\ddot{\phi}-\frac{1}{2}H\frac{df}{d\phi}\dot{\phi}^{3}\Big),
\end{equation}
and
$$P=-V(\phi)\sqrt{1-\dot{\phi}^{2}}+(\xi_{1}+\xi_{2})\big(3H^{2}+2\dot{H}\big)f(\phi)\dot{\phi}^{2}+2(\xi_{1}+\xi_{2})
H\Big(2f(\phi)\dot{\phi}\ddot{\phi}+\frac{df}{d\phi}\dot{\phi}^{3}\Big)$$
\begin{equation}
+(2\xi_{1}+\xi_{2})\Big(f(\phi)\ddot{\phi}^{2}+f(\phi)\dot{\phi}\dddot{\phi}+\frac{5}{2}\frac{df}{d\phi}
\dot{\phi}^{2}\ddot{\phi}+\frac{1}{2}\frac{d^{2}f}{d\phi^{2}}\dot{\phi}^{4}\Big).
\end{equation}
The modified Friedmann equation for the $(0,0)$ component takes the
form,
\begin{equation}
H^{2}=\frac{\kappa^{2}}{3}\Bigg(\frac{V(\phi)}{\sqrt{1-\dot{\phi}^{2}}}+9\xi_{1}
H^{2}f(\phi)\dot{\phi}^{2}+3(2\xi_{1}+\xi_{2})\Big(\dot{H}f(\phi)\dot{\phi}^{2}-H
f(\phi)\dot{\phi}\ddot{\phi}-\frac{1}{2}H\frac{df}{d\phi}\dot{\phi}^{3}\Big)\Bigg).
\end{equation}
Next, we want to study the effects of general non-minimal derivative
couplings on the cosmological evolution of the EoS and see how the
present model can be used to realize a crossing of phantom divide
$\omega=-1$.

\section{Crossing of the $\omega=-1$  with General Non-minimal Derivative Couplings}
We now study the behavior of the equation of state for the present
model. From the definition of EoS  ($\omega=\frac{P}{\rho}$) one can
obtain $P+\rho=(1+\omega)\rho$. Using equations (11) and (12) we
have the following expression,
$$\rho+P=\frac{V(\phi)
\dot{\phi}^{2}}{\sqrt{1-\dot{\phi}^{2}}}+3(4\xi_{1}+\xi_{2})
H^{2}f(\phi)\dot{\phi}^{2}+(8\xi_{1}+5\xi_{2})\dot{H}f(\phi)\dot{\phi}^{2}$$
\begin{equation}
-(2\xi_{1}-\xi_{2})
H\Big(f(\phi)\dot{\phi}\ddot{\phi}+\frac{1}{2}\frac{df}{d\phi}\dot{\phi}^{3}\Big)
+(2\xi_{1}+\xi_{2})\Big(f(\phi)\ddot{\phi}^{2}+f(\phi)\dot{\phi}\dddot{\phi}+\frac{5}{2}\frac{df}{d\phi}
\dot{\phi}^{2}\ddot{\phi}+\frac{1}{2}\frac{d^{2}f}{d\phi^{2}}\dot{\phi}^{4}\Big).
\end{equation}
Since $P+\rho=(1+\omega)\rho$, one needs $\rho+P=0$ when $\omega$
goes to $-1$. Then to check the possibility of the crossing of the
phantom divide line $\omega=-1$, we must explore for conditions
that $\frac{d}{dt}(\rho+P)\neq0$ when $\omega$ crosses over $-1$.\\
Using equation (14) we derive the following equation,
$$\frac{d}{dt}(\rho+P)=\frac{V(\phi)
\dot{\phi}^{3}}{\sqrt{1-\dot{\phi}^{2}}}+\frac{2V(\phi)
\dot{\phi}\ddot{\phi}}{\sqrt{1-\dot{\phi}^{2}}}+\frac{V(\phi)
\dot{\phi}^{3}\ddot{\phi}}{\big(1-\dot{\phi}^{2}\big)^{\frac{3}{2}}}+
3(4\xi_{1}+\xi_{2})\Big(2H^{2}f(\phi)\dot{\phi}\ddot{\phi}$$
$$+H^{2}\frac{df}{d\phi}\dot{\phi}^{3}+2H\dot{H}f(\phi)\dot{\phi}^{2}\Big)+
(8\xi_{1}+5\xi_{2})\Big(2\dot{H}f(\phi)\dot{\phi}\ddot{\phi}+\dot{H}\frac{df}{d\phi}\dot
{\phi}^{3}+\ddot{H}f(\phi)\dot{\phi}^{2}\Big)$$
$$-(2\xi_{1}-\xi_{2}) \Big(\dot{H}f(\phi)\dot{\phi}\ddot{\phi}+H
f(\phi)\big(\dot{\phi}\dddot{\phi}+\ddot{\phi}^{2}\big)+\frac{1}{2}\frac{d
f}{d\phi}\dot{\phi}^{2}\big(5H\ddot{\phi}+\dot{H}\dot{\phi}\big)+\frac{1}{2}
H\frac{d^{2}f}{d\phi^{2}}\dot{\phi}^{4}\Big)$$
\begin{equation}
+(2\xi_{1}+\xi_{2})\Big(
f(\phi)\big(\dot{\phi}\ddddot{\phi}+3\ddot{\phi}\dddot{\phi}\big)+\frac{d
f}{d\phi}\dot{\phi}\big(6\ddot{\phi}^{2}+\frac{7}{2}\dot{\phi}\dddot{\phi}\big)+\frac{9}{2}
\frac{d^{2}f}{d\phi^{2}}\dot{\phi}^{3}\ddot{\phi}+\frac{1}{2}
\frac{d^{3}f}{d\phi^{3}}\dot{\phi}^{5}\Big).
\end{equation}
Now, we mention the following point: if we consider the restriction
(2) then one of the possibilities to have $\rho+P=0$ and
$\frac{d}{dt}(\rho+P)\neq0$ is $\dot{\phi}=0$ [39]. But in the case
of our interest $(2\xi_{1}+\xi_{2}\neq0)$, the condition
$\dot{\phi}=0$ leads to $\rho+P=0$ and $\frac{d}{dt}(\rho+P)=0$ i.e.
the impossibility for having $\omega=-1$ crossing. So, in order to
have crossing of
the phantom divide the only possibility is as follows,\\
$$\dot{\phi}^{2}\Bigg(\frac{V(\phi)
}{\sqrt{1-\dot{\phi}^{2}}}+3(4\xi_{1}+\xi_{2})
H^{2}f(\phi)+(8\xi_{1}+5\xi_{2})\dot{H}f(\phi)-\frac{1}{2}(2\xi_{1}-\xi_{2})
H\frac{df}{d\phi}\dot{\phi}
$$
\begin{equation}
+\frac{1}{2}(2\xi_{1}+\xi_{2})\Big(5\frac{df}{d\phi}
\dot{\phi}^{2}\ddot{\phi}+\frac{d^{2}f}{d\phi^{2}}\dot{\phi}^{4}\Big)\Bigg)=(2\xi_{1}-\xi_{2})
Hf(\phi)\dot{\phi}\ddot{\phi}-(2\xi_{1}+\xi_{2})f(\phi)\big(\ddot{\phi}^{2}+\dot{\phi}\dddot{\phi}\big).
\end{equation}
The above condition simplifies the equation (15) as,
$$\frac{d}{dt}(\rho+P)=\frac{V(\phi)
\dot{\phi}^{3}}{\sqrt{1-\dot{\phi}^{2}}}\Big(1+\frac{\ddot{\phi}}{1-\dot{\phi}^{2}}\Big)
+3(4\xi_{1}+\xi_{2})H^{2}\frac{df}{d\phi}\dot{\phi}^{3}+(8\xi_{1}+5\xi_{2})\ddot{H}f(\phi)\dot{\phi}^{2}$$
$$-(2\xi_{1}-\xi_{2}) \Big(H f(\phi)\big(\dot{\phi}\dddot{\phi}-\ddot{\phi}^{2}\big)+\frac{3}{2}H\frac{d
f}{d\phi}\dot{\phi}^{2}\ddot{\phi}+\frac{1}{2}
H\frac{d^{2}f}{d\phi^{2}}\dot{\phi}^{4}\Big)+(2\xi_{1}+\xi_{2})\Big(
f(\phi)\big(\dot{\phi}\ddddot{\phi}$$
\begin{equation}
+\ddot{\phi}\dddot{\phi}-2\frac{\ddot{\phi}^{3}}{\dot{\phi}}\big)+\frac{d
f}{d\phi}\dot{\phi}\big(\ddot{\phi}^{2}+\frac{7}{2}\dot{\phi}\dddot{\phi}\big)+\frac{7}{2}
\frac{d^{2}f}{d\phi^{2}}\dot{\phi}^{3}\ddot{\phi}+\frac{1}{2}
\frac{d^{3}f}{d\phi^{3}}\dot{\phi}^{5}\Big).
\end{equation}

One can see from (17) that, even if $\ddot{\phi}=0$ and
$\dddot{\phi}=0$, crossing $-1$ can be happen. So, crossing of the
phantom divide in our model can occur in the minimum of the tachyon
potential where one expects $\dot{\phi}\neq 0$ and
$\ddot{\phi}=\dddot{\phi}=0$.\\
This outcome is the same as the result of Ref. [17] where tachyon
field non-minimally coupled to Gauss-Bonnet invariant but in
contrast with the result of Ref. [40], where the authors have added
a term $\phi\Box\phi$, in the square root part of action (1) without
non-minimal derivative coupling terms and concluded that crossing
over $-1$ must takes place before reaching potential to its minimum.
Note that in our model there is no extra term but we have included
non-minimal coupling of tachyon field with its derivative and
curvatures.\\
In summary, it seems that in studying phantom divide crossing
cosmology the non-minimal coupling of tachyon field with its
derivative and the Ricci curvatures has the same effects as coupling
of tachyon to Gauss-Bonnet invariant where crossing over $-1$ can be
happen when tachyon potential reaches its minimum asymptotically.\\
In order to show that our model can realize crossing of $\omega=-1$
more clearly, we choose two specific tachyon potentials and study
evolution of EoS numerically. Figure 1 shows such a numerical
calculations for $V(\phi)=V_{0}e^{-\alpha \phi^{2}}$ with constant
$\alpha$ and for another tachyon potential
$V(\phi)=\frac{V_{0}}{\phi^{2}}$. It has been shown that crossing of
$\omega=-1$ can be realized in our model. Also we have
used the function $f(\phi)=b\phi^{n}$ with constants $b$ and $n$.\\

\begin{figure}[htp]
\begin{center}
\includegraphics{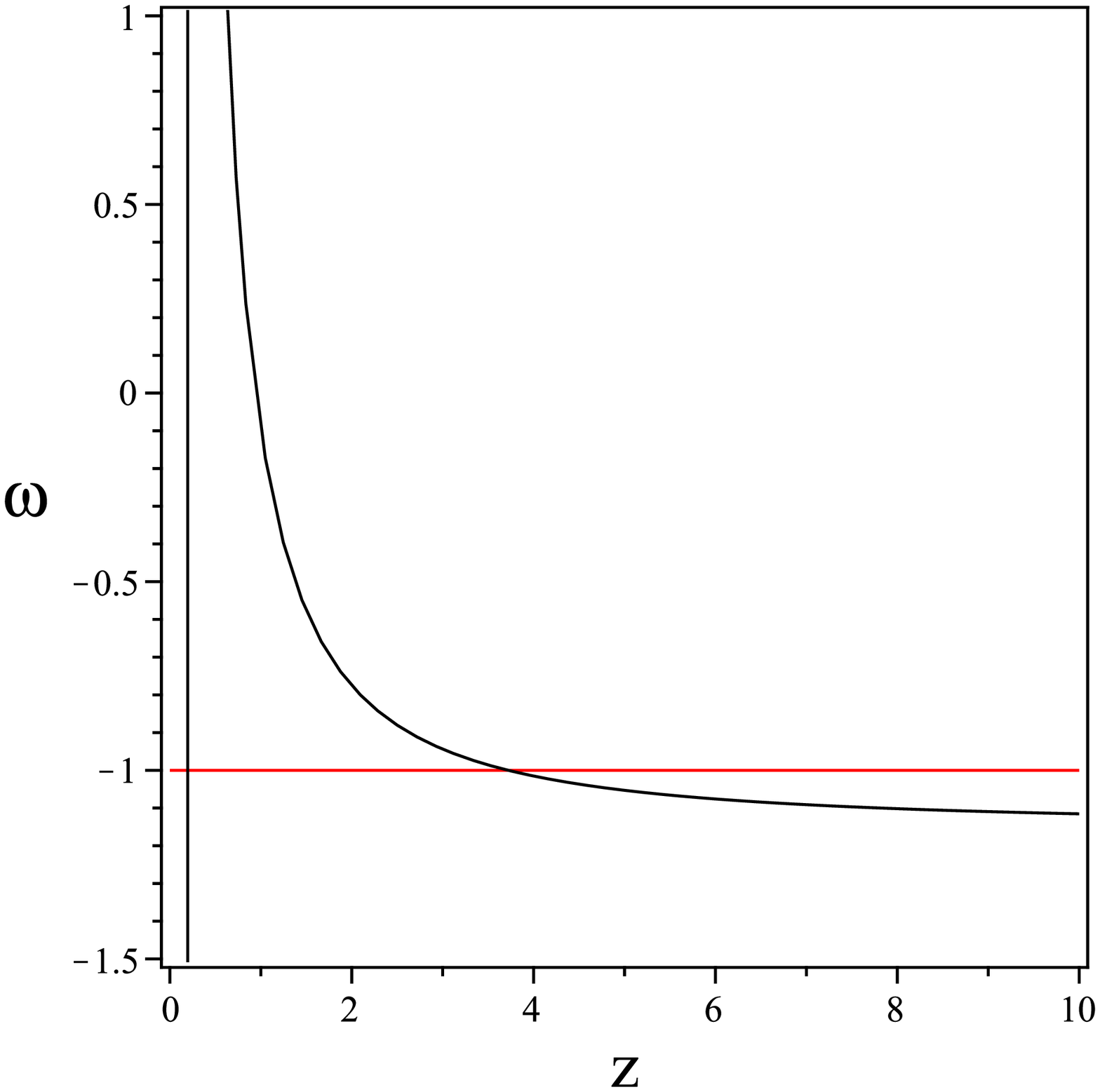} \vspace{8.5cm}\includegraphics{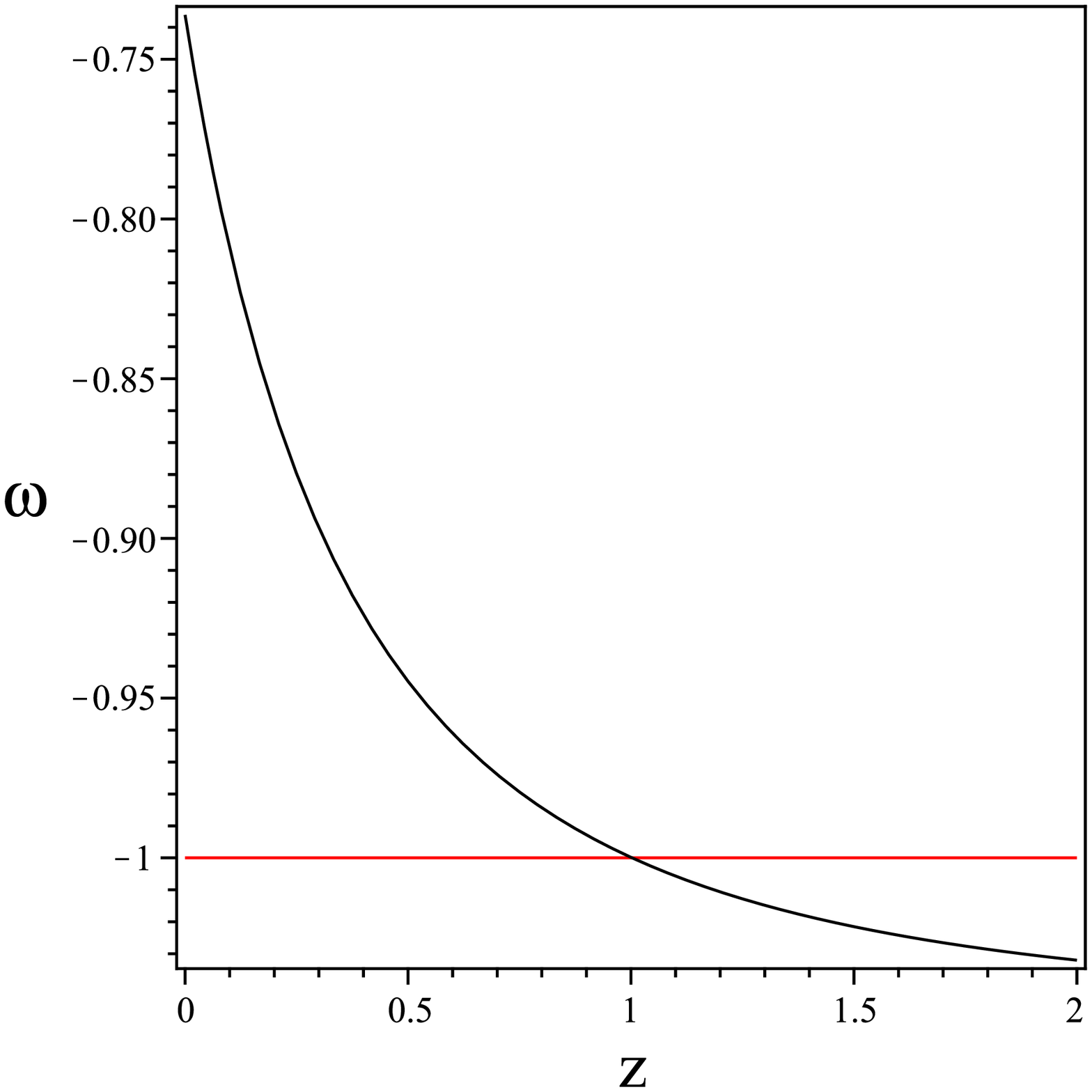}
\end{center}
 \caption{\small { The plots of EoS versus redshift $z$, (left for the potential
$V(\phi)=V_{0}e^{-\alpha \phi^{2}}$ and right for the potential
$V(\phi)=\frac{V_{0}}{\phi^{2}}$) , $\phi=\phi_{0}t$,
$f(\phi)=b\phi^{n}$ and $H=\frac{h_{0}}{t}$, (with
$\xi_{1}=\xi_{2}=10$, $b=1$, $n=5$, $V_{0}=4$, $h_{0}=100$,
$\phi_{0}=0.5$ and $\alpha=5$).}}
\end{figure}

Now we discuss on the stability of the model. The sound speed
expresses the phase velocity of the inhomogeneous perturbations of
the tachyon field. To achieve the classical stability, we must have
$C_{s}^{2}\geq0$, where
$$C_{s}^{2}=\frac{P'}{\rho'}$$
\begin{equation}
=\frac{\frac{1}{2}\frac{V(\phi)}{\sqrt{1-\dot{\phi}^{2}}}+(\xi_{1}+\xi_{2})\Big((3H^{2}+2\dot{H})f+2
H\big(f\frac{\ddot{\phi}}{\dot{\phi}}+\frac{3}{2}\frac{df}{d\phi}\dot{\phi}\big)\Big)+(2\xi_{1}+\xi_{2})\big(\frac{1}{2}f
\frac{\dddot{\phi}}{\dot{\phi}}+\frac{5}{2}\frac{df}{d\phi}\ddot{\phi}+\frac{d^{2}f}{d\phi^{2}}\dot{\phi}^{2}\big)}
{\frac{1}{2}\frac{V(\phi)}{(1-\dot{\phi}^{2})^{\frac{3}{2}}}+9\xi_{1}H^{2}f+3(2\xi_{1}+\xi_{2})\big(\dot{H}f-\frac{1}{2}H
f\frac{\ddot{\phi}}{\dot{\phi}}-\frac{3}{4}H\frac{df}{d\phi}\dot{\phi}\big)},
\end{equation}
where a prim denotes derivative with respect to $\dot{\phi}^{2}$.\\
In figure 2, we have plotted the $C_{s}^{2}$ for the models
considered in this paper for the numerical calculations. From this
figure, we can see that the sound speed parameter is positive
throughout the phantom divide crossing phase.
\begin{figure}[htp]
\begin{center}
\includegraphics{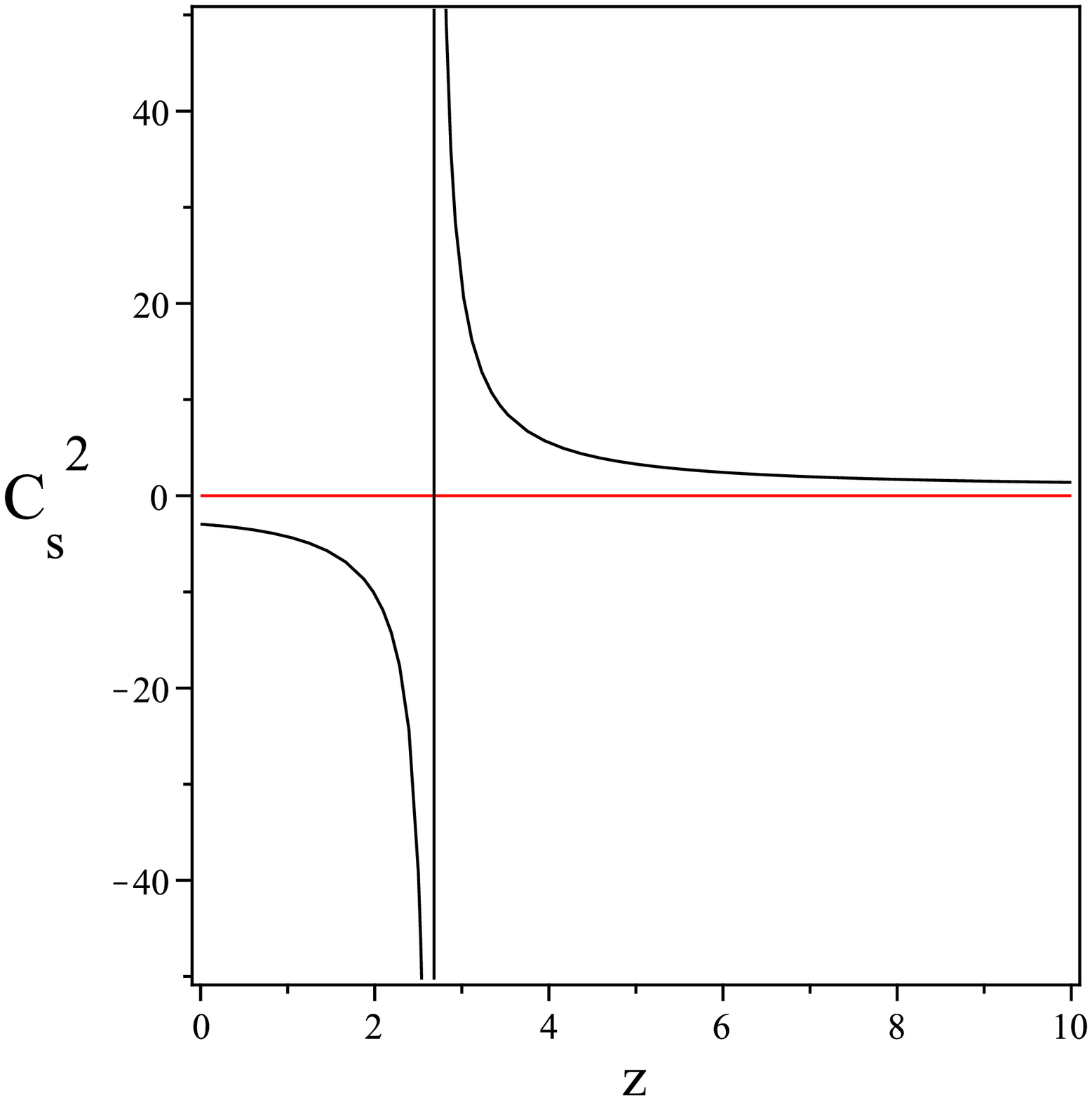} \vspace{7cm}\includegraphics{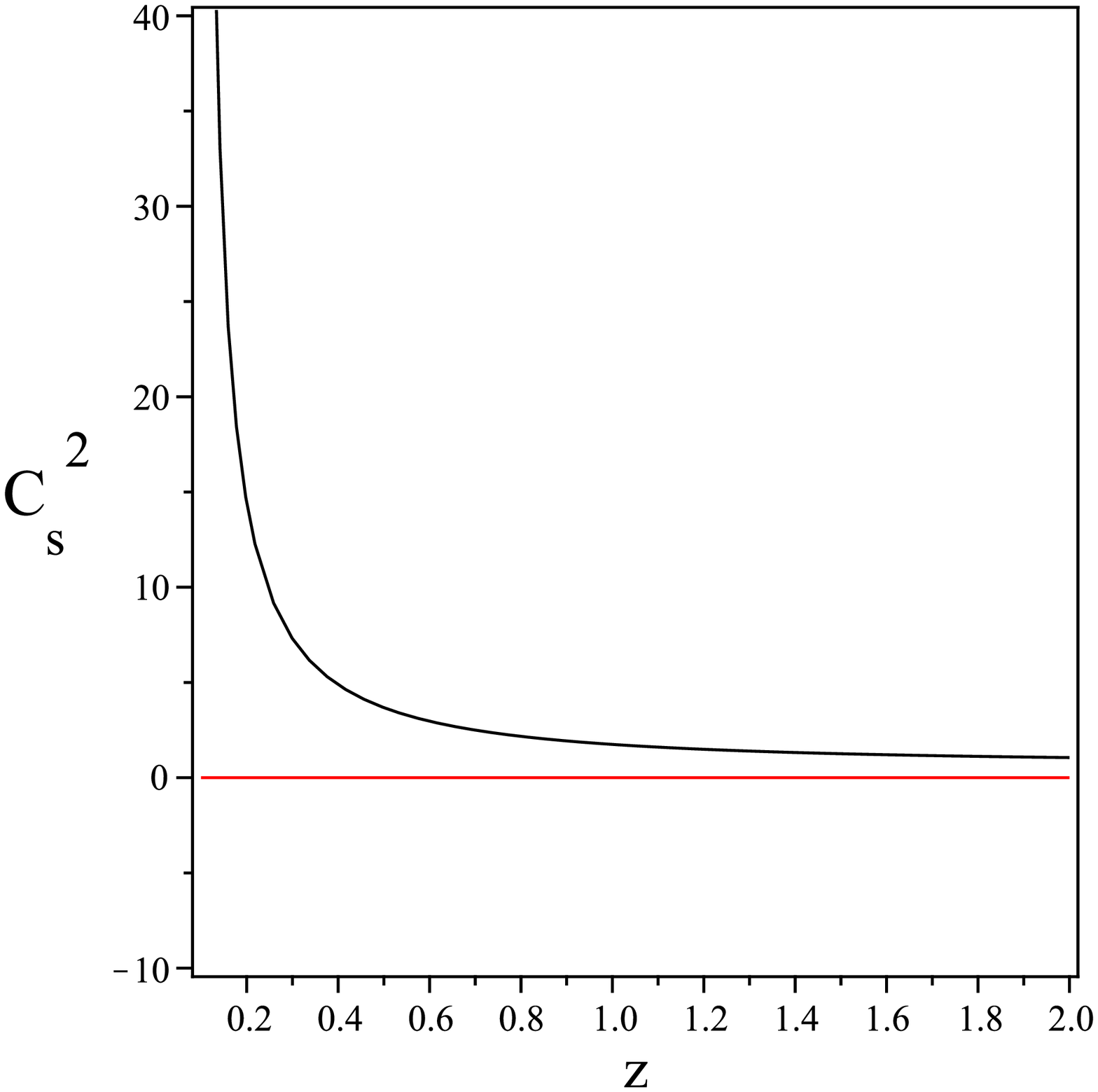}
\end{center}
\caption{\small { The plots of the sound speeds versus redshift $z$,
(left for the potential $V(\phi)=V_{0}e^{-\alpha \phi^{2}}$ and
right for the potential $V(\phi)=\frac{V_{0}}{\phi^{2}}$) ,
$\phi=\phi_{0}t$, $f(\phi)=b\phi^{n}$ and $H=\frac{h_{0}}{t}$, (with
$\xi_{1}=\xi_{2}=10$, $b=1$, $n=5$, $V_{0}=4$, $h_{0}=100$,
$\phi_{0}=0.5$ and $\alpha=5$).}}
\end{figure}

\section{Conclusion}
In order to solve cosmological problems and because the lake of our
knowledge, for instance to determine what could be the best
candidate for DE to explain the accelerated  expansion of the
universe, the cosmologists try to approach to best results as
precise as they can by considering all the possibilities they
have.\\
The two most reliable and robust SnIa datasets existing at present
are the Gold dataset [1] and the Supernova Legacy Survey (SNLS) [43]
dataset. The Gold dataset compiled by Riess et. al. is a set of
supernova data from various sources analyzed in a consistent and
robust manner with reduced calibration errors arising from
systematics. It contains $143$ points from previously published data
plus $14$ points with $z > 1$ discovered recently with the HST. The
SNLS is a $5$-year survey of SnIa with $z < 1$. The published first
year SNLS dataset consists of $44$ previously published nearby SnIa
with $0.015 < z < 0.125$ plus $73$ distant SnIa $0.15 < z < 1$. The
following comments can be made for phantom divide crossing based on
the cosmological data [44]: The Gold dataset mildly favors
dynamically evolving dark energy crossing the phantom divide at $z
\simeq 0.2$ over the cosmological constant while the SNLS does not.
Dark energy probes other than SnIa that include the CMB, BAO,
Clusters Baryon Fraction and growth rate of perturbations mildly
favor crossing of the phantom divide for low values of $\Omega_{0m}$
($\Omega_{0m} \leq ­ 0.25$).\\
Within the different candidates to play the role of the DE, the
quintom model, has emerged as a possible model with EoS across $-1$.
In this paper, we have proposed a model of dark energy with
non-minimally kinetic coupled scalar field, where the kinetic term
is not only coupled to itself through the function $f(\phi)$, but to
the Ricci scalar and the Ricci tensor. We have studied cosmological
evolution of EoS in this setup where tachyon field played the role
of scalar field. We considered the non-minimal kinetic couplings,
without the restriction on the coupling constants $\xi_{1}$ and
$\xi_{2}$ namely equation (2) and obtained the condition required
for phantom divide crossing as equation (16). It has been shown that
the $\omega=-1$ crossing can be realized even if the potential goes
to its minimum asymptotically and this result is the same as that in [17].\\
Using the numerical methods we showed that the crossing of phantom
divide occur for special potentials and coupling function. It may be
interesting to consider different potentials and coupling
functions in this setup.\\

\end{document}